\documentclass[prl,twocolumn,showpacs]{revtex4}

\input{epsf}

\newcommand{\bq}{\begin{equation}}
\newcommand{\ee}{\end{equation}}

\begin{document}

\draft

\begin{abstract}
This note has no new results and is therefore not intended to be
submitted to a "research" journal in the foreseeable future, but
to be available to the numerous individuals who are interested in
this issue. Several of those have approached the author for his
opinion, which is summarized here in a hopefully pedagogical
manner, for convenience. It is demonstrated, using essentially
only energy conservation and elementary quantum mechanics, that
true decoherence by a normal environment approaching the
zero-temperature limit is impossible for a test particle which can
not give or lose energy. Prime examples are: Bragg scattering, the
M\"ossbauer effect and related phenomena at zero temperature, as
well as quantum corrections for the transport of conduction
electrons in solids. The last example is valid within the
scattering formulation for the transport. Similar statements apply
also to  interference properties in equilibrium.

\end{abstract}

\title{ Elementary explanation of the inexistence of decoherence at zero
temperature for systems with purely elastic scattering}

\author{Yoseph Imry}

\address{Dept. of Condensed-Matter Physics,\\ the Weizmann Institute of science,\\ Rehovot 76100, Israel}

\pacs{ 73.23.Hk, 73.20.Dx ,72.15.Qm, 73.21.La}

\date{\today}
\maketitle

\section{I. Introduction}

What diminishes the interference of, say, two waves (see Eq.
\ref{TwoWave} below) is an interesting fundamental question, some
aspects of which are, surprisingly,  still being debated. This
process is called "dephasing" or "decoherence" \cite{footnote1}.
Decoherence occurs due to an interaction of the interfering entity
(henceforth referred to as "the particle") with the degrees of
freedom (dubbed "the environment", examples for which are lattice
vibrations, E-M fields, interactions with  other particles, etc.)
which are not measured in the interference experiment. More
specifically, this interaction will eliminate the interference if
and only if the two (in the simplest case) partial waves leave the
environment in orthogonal states. The diminishing of the
interference can then be described as due to either tracing over
the environment's states or to the random fields exerted by the
fluctuating environment on the particle. These two descriptions
are exactly equivalent \cite{FV,SAI}. After almost twenty years of
research on Mesoscopic Physics \cite{book}, it is now fully and
universally accepted that purely {\em elastic} scattering, for
example by static defects in solids, can modify the interference
terms but does not cause decoherence. An extreme case occurs if
one of the two interference paths is blocked. The interference
obviously vanishes (as will a half of the classical contribution).
But this is not decoherence. Likewise, one can diminish the
interference by various averaging processes (For example, via a
broad enough distribution of incident particle energies or
wavelengths) but, again, this is not proper decoherence.

The issue of whether the zero point modes of an environment can
dephase  \cite{footnote1} a test particle which can not lose
energy \cite{book}, is still current in modern literature. An
answer to the affirmative has been suggested theoretically in the
case of the coupling of a conduction-electron in a solid to
lattice vibrations (phonons) 14 years ago \cite{Kumar} and
immediately refuted vigorously \cite{Schm-Berg}. Interest in this
problem has resurfaced due to experiments by Mohanty et al
\cite{Webb} which determine the dephasing rate of conduction
electrons by weak-localization magnetoconductance \cite{Berg}.
According to these rather careful experiments, the dephasing rate
of a conduction electron does not vanish when the temperature $T
\rightarrow 0$, but rather goes to a finite limit which is then
interpreted as due to the coupling with the zero-point
electromagnetic fluctuations in the solid. For conductors, these
fluctuations are mainly due to the fields produced by the other
conduction electrons, hence this decoherence can also be described
as due to electron-electron interactions. This dephasing rate has
been calculated before \cite{AAK} and found to vanish at the zero
temperature limit. An apparent contradiction between theory and
experiment results.

This contradiction is not just with a specific model calculation.
As mentioned above, decoherence is very generally understood as
what may be called a "which path" detection by the "environment",
effected by the exchange of an excitation (i.e. an inelastic
process) \cite{SAI} between the particle and the environment. Such
excitation exchange processes can not occur when both the
environment and the test particle are at zero temperature and in
the linear transport regime (where the conduction electron has an
excess energy which tends to zero) \cite{Zvi,footnote2}. No
excitation can be exchanged if neither the particle nor the
environment can lose energy. It is assumed that, as is usually the
case, the environment does not have a large degeneracy of the
ground state. The above argument that no excitation can be
exchanged between the test-particle and the environment can
actually be formulated in terms of an integral over products of
certain correlation functions, which can rather generally be
proven to vanish as $T \rightarrow 0$ \cite{CI,isqm,IFS,book}. The
only exceptions being an environment having a pathologically
increasing density of states at low energies and the
above-mentioned case where the environment has a large ground
state degeneracy. A  viable model for the latter are uncompensated
and isolated magnetic impurities at zero magnetic field.

However, the experimenters \cite{Webb} have taken great pains to
eliminate the effect of magnetic impurities and have presented
serious arguments as to why that was not an important effect in
their samples. An assumed high enough density of other low-energy
modes, such as two-level systems (TLS) was also shown \cite{IFS}
to be able to account, in principle, for the observed anomalies.
Such low-energy modes might depend on the metallurgy of a given
sample. Other experiments \cite{kha} have reported a vanishing
dephasing rate as $T \rightarrow 0$ in a different material. Later
experiments \cite{Norm} have in fact demonstrated that the $T
\rightarrow 0$ anomaly seems to depend on sample preparation. Thus
some lattice defects such as the TLS might be relevant.
Uncompensated magnetic impurities were also implicated in Ref.
\cite{Norm} but that possibility was argued to have been negated
for the experiments of Ref.\cite{Webb}. Here we certainly do not
attempt to pass judgement on the experimental controversy.
Furthermore, as indicated by some of the early experiments
\cite{Berg} one must use extremely small measurement currents to
be in the linear transport regime. Ovadyahu \cite{Ov00} studied
this issue very carefully. His findings can be summarized as
showing that it may be possible to eliminate the anomalous
decoherence by using extra-small probing currents. However, there
is a large range of currents that do {\em not} heat the electrons,
as found in Ref. \cite{Webb}, but appear nevertheless  not to be
small enough for the transport to be in the truly linear regime.
Finding the mechanism leading to this, which probably also
involves some special low-frequency modes \cite{Ov00}, still
presents an important unsolved problem in  low-temperature metal
Physics. Therefore, we again refrain from offering a verdict on
what the final answer provided by all the experiments will be and
we trust that that will be cleared up soon. From now on we confine
ourselves to the, still hotly debated, theoretical question only,
which ought to be decidable by applying  known principles
correctly.

That question has become controversial as well, due to reports of
calculations \cite{zaikin} which claimed to have produced a finite
dephasing rate as $T \rightarrow 0$, in contradiction with refs.
\cite{AAK,SAI} and with the general arguments summarized above.
Refs. \cite{ale,alt} have disagreed strongly with these
calculations. To which criticism the authors \cite{zaikin} have
rebutted. The theoretical controversy is still going on ( e.g.
\cite{vonDelft}, \cite{Schoen} and \cite{KB}). Unlike the
experimental situation, experience has shown that questions such
as "who has calculated the correct diagrams correctly" tend to
linger and produce unnecessary controversy. This is compounded by
further reports of the weakening of quantum interference, even
truly at equilibrium, at $T \rightarrow 0$, by a coupling to the
environment \cite{But}. It has been claimed that "Vacuum
fluctuations are a source of irreversibility and can decohere an
otherwise coherent process". That statement is of course trivially
valid (if certain matrix-elements do not vanish) for a system that
can lose energy to the vacuum fluctuations. However, that does
{\em not} include an electron just on the Fermi-level. The
persistent current in a mesoscopic  ring was found \cite{But} to
be weakened at $T \rightarrow 0$ by coupling to harmonic
oscillators in their ground state. Models invoking such
oscillators with a specific coupling and density of states (DOS),
chosen  to mimic the dissipation \cite{Nyquist,FV,Zwan,CL} can be
used to describe a viscous resistance to the electron's motion
(although, at best, important modifications, such as a spatial
distribution of these oscillators might be necessary
\cite{Doron,CI}).

Thus, the theoretical situation needs clarification, independently
of the final experimental  verdict. We shall attempt in this very
informal note to clarify the possible misunderstandings that can
lead to the belief in the $T \rightarrow 0$ decoherence. We shall
show that some of the confusion may be due to interpreting a
reduction of interference as due to dephasing, when it really  is
something else. To explain that, we shall first treat in section
II an elementary model, really a rehash of the arguments of Ref.
\cite{Schm-Berg}, which illustrates the whole issue very simply.

\section  {II. A simple model: two-wave interference with vibrating
scatterers } \label{simple}

Let us first imagine two elastic, rigid, point scatterers placed
at points $\vec x_1$ and $\vec x_2$, separated from each other by
a vector $\vec d$ along the x axis. A particle wave with wave
vector $\vec k_0$ impinges upon these scatterers. We look at the
(elastic) scattering into the state with a wavevector $\vec k$. We
denote the momentum-transfer vector by $\vec K = \vec k_0 - \vec
k$. If the scattering amplitude from each of the scatterers is
$A_K$, the (lowest order) scattering probability from the system
will be proportional to

\bq \label{TwoWave} S_K = 2|A_K|^2 + 2 {\rm Re} |A_K|^2 \exp(i
\vec K \cdot \vec d )\equiv S_K^{cl} + S_K^{qu}, \ee

\noindent where the first and second terms are, in obvious
notation, the classical and  the interference contributions,
bearing a strong similarity to the case of diffraction by a double
slit. $\vec K \cdot \vec d$ is the phase shift due to the
difference in the optical paths between the waves scattered by the
two scatterers. Somewhat similarly to the Bohr discussion of
diffraction from  two fluctuating slits, we now let each scatterer
be bound by a parabolic potential, so that the frequency of the
motion in the x-direction of  is $\omega_0$. The case of interest
to us here is when the scatterers are at zero temperature, but the
generalization is obvious. We write $\hat x_i = X_i+ {\hat u_i}$,
$i = 1,2$. Each $\hat x_i$ performs zero-point fluctuations (whose
coordinate operator is $\hat u_i$) around its equilibrium position
$X_i = \langle x_i \rangle$. To obtain the new scattering pattern
we must average the scattered $intensity$ over the wavefunction of
the scatterers. That wavefunction is a product of two gaussians,
of $u_1$ and $u_2$. The standard deviation of each is $\sigma$,
given by $\sigma ^2 = \frac{\hbar}{2M \omega_0}$, where $M$ is the
mass of the scatterer. Using the fact that for a gaussian
distribution of $v$ with an average $\langle v \rangle$ and a
variance $\langle \Delta v^2 \rangle$,

\bq \label{Gauss} \langle \exp(iv) \rangle= \exp(i \langle v
\rangle - \langle \Delta v^2  \rangle /2) \ee

\noindent (a similar result is valid also in the quantum case
\cite{Kittel,Messiah}), we obtain

\bq \label{av} S_K = \langle S_K^{cl}\rangle + 2 {\rm Re} |A_K|^2
\exp(i \vec K \cdot \vec d) \exp(-2W), \ee

\noindent where $2W$ is the well-known Debye-Waller factor

\bq \label{DW} 2W = K_x^2 \sigma ^2, \ee

\noindent obtained by writing $\langle (u_1 - u_2)^2 \rangle = 2
\sigma ^2$, since the $u'$s are independent random variables with
zero mean value  each.

Thus, the quantum interference term is indeed reduced by the
zero-point motion. One may (as several researchers have done) now
jump to the {\em wrong} conclusion that the zero-point motion
indeed reduces the quantum interference term without apparently
affecting the classical terms. In other words, one would write:

\bq \label{wrong} \langle S_K^{cl}\rangle = 2|A_K|^2. \ee

\noindent This would imply decoherence by the zero-point motion!
The ({\em superficially} valid only) reason for the above mistaken
conclusion is that in the classical terms one has $\langle e^{iK_x
u} e^{-iK_x u}\rangle = 1$.

The fallacy in the above naive thinking is  in the application of
the last seemingly innocuous  equality, Eq. \ref{wrong}. Because
it automatically includes all  the  inelastic scattering
\cite{stat}, which should {\it not} appear. This fact has been
known for at least 45 years \cite{Zemach,VH} and has been
reinforced when the M\"ossbauer effect was discovered and
understood (see, for example, \cite{Lipkin,Kittel}). Next, we will
explain this issue.

Let us review the derivation of the inelastic nature \cite{VH} of
part of the above expressions. The exact elastic scattering
amplitude $A_K$ of the isolated static scatterer is first
parameterized in terms of a suitably defined pseudopotential
$V(r)$, whose Fourier transform is $V_K$. As if anticipating  the
current sophisticated red-herring type discussions of the
perturbative vs the "nonperturbative" nature of the theory, Fermi
in 1936 \cite{Fermi} defined a pseudopotential $V(r)$, so that the
lowest-order Born approximation scattering amplitude due to it
(which is proportional to $V_K$) equals the actual, correct,
$A_K$. Therefore,  using the Born approximation with the
pseudopotential, already implies taking into account all the
multiple scattering by {\em the same} \cite {multip} scatterer.
Within this  picture, van Hove considered in 1954 the (in general)
inelastic Born approximation scattering, which for our system  is
proportional to

\begin{eqnarray} \label{VH1}  S(K,\omega)= |A_K|^2\sum_f |\langle f| [\exp (i
K_x \hat x_1) +  \nonumber \\ \exp (i K_x \hat x_2)]| g\rangle |^2
\delta (\hbar \omega + E_f- E_g)  \end{eqnarray}

\noindent where $ \hbar \omega$ is the energy, $(\hbar^2/2m)
(k_0^2 - k^2)$, lost by the scattered  particle and $E_f$ and
$E_g$ are the energies of the final state ($|f\rangle$) and
initial one ($|g\rangle$) of the oscillators. The $\delta$
function simply gives total energy conservation in the process.
Since the scatterers are at $T \rightarrow 0$, the only possible
initial state is the ground state, $|g\rangle$, of the
two-oscillator system.

A slightly  more advanced treatment of Eq. \ref {VH1} will be
given in the next section. Right now, we simply enforce the
condition of no energy exchange in the scattering. We remind the
reader that for electrons in metals, this most important condition
follows from the fact that neither the oscillators (which are at
their ground states) nor the conduction electron at the Fermi
level have any states to go down to. This is valid {\em for an
arbitrary, interacting, system} because there are no states below
the ground state. In the noninteracting fermion picture, this
would follow formally, from the Fermi-Pauli factors ($n_g (1 -
n_f)$) that should then be included in Eq. \ref{VH1}.  This
enforces the choice of only the term with $f = g$ in the sum. As a
result, the $\delta (\omega)$ part of $S(K, \omega)$, which is the
elastic scattering cross section and the only surviving part of
$S(K, \omega)$, is given here by

\begin{eqnarray} S_K \equiv \int d\omega S^{el}(K, \omega) = |A_K|^2 |\langle
g| [\exp (i K_x \hat x_1) + \nonumber\\ \exp (i K_x \hat x_2)]|
g\rangle |^2 \label{VH2} \end{eqnarray}

\noindent We again have a classical term which is a sum of a term
due to $x_1$ and one due to $x_2$, and the interference term which
is the ground-state average of $2 {\rm Re} |A_K|^2 \exp(i  K_x
\hat x_1) \exp(-i  K_x  \hat x_2)$, yielding the Debye-Waller
factor as before. The whole point is now, that each classical term
is given by $|A_K \langle g|\exp(i K_x x_i)|g\rangle |^2, ~~i =
1,2$. Therefore, the classical terms too are reduced by the {\em
same Debye-Waller factors} as the interference term. Thus, again,
it is not that the classical terms stay unchanged and the
interference is decreased, as would be the case for a true
decoherence, but {\em both} the non-interfering classical part and
the interference, "quantum correction" part are renormalized in
the same fashion. Therefore, the reduction of the interference
part by the coupling to the environment {\em has nothing to do
with decoherence}. Looking at the average of the classical terms,
it is instructive to point out that the reduction of the
scattering in a distributed scatterer is due to the addition of
the scattering amplitudes from the various positions occupied by
that scatterer. These amplitudes add coherently because the
scattering is elastic. This amplitude addition is familiar from
diffraction theory in optics or from antenna theory.

\section{III. A  more advanced treatment}

For a fuller treatment, van Hove \cite{VH} started by writing Eq.
\ref{VH1} for N scatterers. The term in the square brackets is
just $\sum_{i=1,N} \exp(i{\vec K}\cdot \hat {\vec x}_i)$, which is
just the Fourier transform, $n_K$ of the particle density operator
$ \hat n(\vec x) \equiv \sum_{i=1,N} \delta( {\vec x} -\hat {\vec
x}_i) $. He then used a series of simple but ingenious
manipulations (Fourier-representing the energy $\delta$ function,
inserting a complete set of states and using the Heisenberg
representation for the $n_K$) to write $S(K,\omega)$ as the
Fourier transform of the temporal correlation function, $\langle
n_{-K}(0) n_{K}(t) \rangle$. For phonon systems, where ${\vec
u}_i$ (the fluctuation of ${\vec x}_i$ from its average position
in the lattice) is a linear superposition of harmonic oscillators,
the correlators relevant for inelastic neutron scattering and
treated extensively in that literature in the late 50's (of the
last century) \cite{Kittel}), are therefore

\bq \exp(-J_{ij}(t)) = \langle \exp(i{\vec K}\cdot \hat {\vec
x}_i(t)) \exp(-i{\vec K}\cdot \hat {\vec x}_j(0))\rangle \ee

\noindent The {\em elastic} part of the scattering is given by the
$t \rightarrow \infty$ limit (or time-independent) part of this,
since the Fourier transform of a constant is a $\delta$ function.
In the modern so-called $P(E)$ \cite{PE} theory of phase
fluctuations, similar calculations have been done and similar
functions appear with harmonic oscillators taken to represent the
lossy environment. Looking at the interference between two paths,
one again sees immediately that the two-path interference term is
affected, but that at the same time the single-path terms are
affected in  the same fashion. This completes our demonstration
that zero-point fluctuations do not decohere. Some details will
still be presented below.

It is perhaps useful to point out that the mistake in neglecting
the Debye-Waller factors in the first terms of Eqs. 1 and 3 occurs
because there one looks at the equal-time correlators. These are
related to the integrals of $S(K, \omega)$ over all frequencies
\cite{stat}. Now, the inelastic parts of $S(K, \omega)$ have to be
discarded, as discussed above.
The $t = 0$ correlator (leading to Eq. \ref{wrong}), obtained by
neglecting this important constraint, is very different from the
relevant $t \rightarrow \infty$  limit of the correlator (leading
to Eq. \ref{VH2}), which is related to the {\em elastic
scattering}. We are now ready to discuss the question of what
happens when the $t \rightarrow \infty$ limit of the above
correlators vanishes. The answer depends on the physical situation
considered. In  the scattering problem, the vanishing of the
Debye-Waller factor means that there is no strict Bragg-type
scattering or M\"ossbauer effect (but interesting things can still
happen, see for example Refs. \cite{GI,IG}). If the inelastic
scattering is blocked, no scattering remains and the particle can
only go forward "unscathed" through the system (remaining
unscattered). However, other situations can exist. One can have a
vibrating double-slit system that will stop transmitting
altogether. An electron in a solid with phonons can also localize.
A sizable increase of the electron's effective mass was  found by
Holstein \cite{Holstein} in the small-polaron problem (which is
also related to the issues discussed here). Later,  Schmid
\cite{Schmid} found that for the phonon problem similar to the one
discussed here, with a suitable "ohmic" spectrum, this mass
renormalization is so severe that the mass diverges for a strong
enough coupling, and the particle becomes localized. This effect
appeared later, in the context of what has been  called
"Macroscopic Quantum Coherence" \cite{Legg}, for mesoscopic
SQUIDS. It should perhaps be reemphasized that this mass
renormalization and possible ensuing localization are due to an
elastic "orthogonalization catastrophe" and {\em not} to
decoherence \cite{Stern}. We shall  discuss this further in the
next section.

It has become customary to describe  the finite resistance of an
electron in a real metal as due to coupling with the
Nyquist--Feynmann-Vernon--Zwanzig--Caldeira-Leggett oscillators
\cite{expl}. We put aside, for the time being, the serious
question of how well can a resistance given physically when $T
\rightarrow 0 $ by (mainly) elastic impurity scattering, be
described by inelastic (and therefore phase-breaking) processes.
Within this model the  phase decay for an electron coupled to a
dissipative bath is described by the temporal correlator $\langle
\exp(i\hat \phi(t)) \exp(-i\hat \phi(0))\rangle$. It is governed
\cite{Im} when $T \rightarrow 0$ by

\bq \exp[- const \int_0^\infty d\omega \eta(\omega)\,
(\frac{1-\cos\omega t}{\omega})], \label{DW1} \ee

\noindent where $\eta(\omega)$ is the dissipation constant,
assumed to have a finite $\omega \rightarrow 0$ limit. We
emphasize that this is the naive result, obtained by neglecting
the fact that inelastic processes are absent. To set up an analogy
with a real lattice dynamics problem we note that the "self"
correlator for the single-particle density $\langle \exp(iK\hat
x_i(t)) \exp(-iK\hat x_i(0))\rangle = \exp(-J_{ii}(t))$ at $T = 0$
for a 1D Debye lattice is also (see e.g. Ref. \cite{IG}) decaying
(see \cite{Im}) via $\exp[(- const \int_0^{\omega_D} d\omega
(\frac{1-\cos\omega t}{\omega})]$, \cite{Im}. Thus, these two
problems are mathematically similar. Both have, when $T
\rightarrow 0$, the same logarithmic long-time behavior of the
exponent, which becomes a power-law decay of the correlator.
However, in the latter case, these zero point-effects are very
physical -- for example, an atom emitting radiation can produce
phonons (in the non-M\"ossbauer channels) in its recoil. However,
as we explained, the electron on the Fermi level can not decay and
produce an excitation. Its inelastic channels are blocked. The
amplitude for the particle to go between two points is the sum of
the amplitudes of a huge number, $\cal N$, of paths. The
probability has two types of terms: $\cal N$ classical ones (the
absolute value squared of the amplitude of a single path) and
${\cal N}^2 / 2$ interference terms (twice the real part of the
product of a path amplitude with the complex conjugate of the
amplitude of a different path). Once the constraint of no energy
transfer is imposed, there is again no major difference between
the (Debye-Waller type) reduction factors for the classical and
the interference terms. Hence there is no intrinsic decoherence in
the Physics of a particle coupled to a normal bath at $T
\rightarrow 0$. Unusual low-energy properties of the bath might of
course still give anomalous decoherence at low temperatures. Such
anomalies do not exist in the theoretical models on which the
present controversy rages
\cite{zaikin,ale,alt,Schoen,vonDelft,KB}.

\section{IV. Quantum effects in closed rings, "antidecoherence"???}
Besides real decoherence, there are other ways to reduce quantum
interference effects.  Averaging over the energy of the incoming
particle is often confused with decoherence, but it should not
\cite{book}. Using a variable energy filter one may recover the
many interference patterns belonging to different energies in the
beam. This is impossible for decoherence via "which  path"
identification. A rather aggressive way to kill quantum
interference is, for example, to simply block (at least) one of
the two interfering paths. Alternatively, if the Fermi-surface
electron interacts with the $T \rightarrow 0$ bath only when it
goes via one of the interfering paths , the amplitude of that path
will be reduced by a Debye-Waller type factor similar to Eq.
\ref{DW1} (see \cite{Im}),

\bq \exp[- const \int_0^\infty d\omega \lambda(\omega)\,
(\frac{1-\cos\omega t}{\omega})], \label{DW2} \ee

\noindent where $\lambda(\omega)$  contains the coupling constants
and the DOS of the oscillators at frequency $\omega$. The limit of
effectively "cutting"  the path is achieved when this factor
becomes very small. A similar reduction will be achieved if only
one of the scatterers in the simple model of section II, is
allowed to fluctuate. If the two paths are cut or severely
weakened, a decrease of both the interference and the classical
terms is obtained. It is hoped that the reader already understands
very well that this reduction is distinct from decoherence.

One may now consider a case where, in a sense, there are {\em no}
classical terms. The sensitivity of the energy levels of an
electron on an Aharonov-Bohm (AB) ring to the AB flux is a good
example. There is no such sensitivity in the classical case. This
sensitivity (which determines for example the persistent current
through the ring, see, for example, \cite{book}) follows from the
phase shift experienced by  the electron going around the ring
enclosing the flux. This AB phase shift along such paths
determines the flux sensitivity of, say, the energy levels and
thence the persistent current. Obviously, if decoherence is strong
enough, in the sense that the ring's circumference $L$ is much
larger than the electron's coherence length $L_{\phi}$, these
purely quantum effects are exponentially reduced.

In the Holstein polaron model \cite{Holstein}, the effective mass
of the electron increases by the coupling with the zero-point
environment. Imagine putting such an  electron on an AB ring. The
flux-sensitivity of the energy levels is proportional to the
inverse mass. It will be reduced by the increase of the latter. In
certain cases \cite{Schmid,Legg} the effective mass diverges (see
the previous section) and the flux sensitivity is altogether
eliminated. However, describing this electron with the measured,
renormalized, effective mass, will eliminate any need to even
think about decoherence.

Likewise, the coupling of a Fermi-surface electron to a $T
\rightarrow 0$ oscillator bath will reduce the amplitude of its
flux-encircling paths by the  Debye-Waller type factor, Eq.
\ref{DW2}. Other paths, not encircling the ring, will experience
the corresponding reductions as well, which will just renormalize
the physics of the ring.  The magnitude of the AB oscillations
will be suppressed by such factors for the relevant paths. This is
the effect found in Ref. \cite{But}. Quantum interference is
indeed reduced, but this is not decoherence.

We now prove the incorrectness of identifying the Debye-Waller
reduction of interference as decoherence by {\it reductio ad
absurdum}. It is easy to devise couplings to an external bath
which {\em enhance} the effects of quantum interference. In a
diffusive real mesoscopic ring, the persistent current is well
known \cite{book} to increase with the elastic mean free path
$\ell$. The physics of the conduction electrons is governed by two
Debye-Waller renormalizations, that of the scattering by the
impurities (\ref{DW}), due to lattice vibrations and that of the
electrons (\ref{DW1}) due basically to the electron-electron
interactions. Coming from different interactions, these two
renormalizations are not strongly dependent (see below) on each
other. Suppose, for simplicity, that all relevant atomic masses
are light enough so that the former is more important for the
renormalized conductivity than the latter. As found in section II,
the elastic impurity scattering rate is proportional to the
Debye-Waller factor (for further discussions of this, including
the  more complicated case of finite temperatures, see for example
Refs. \cite{Kagan,Misha}). Thus, the magnitude of the quantum
interference for $T\rightarrow 0$ will {\em increase} with the
decreasing vibrational Debye-Waller factor (\ref{DW}) of the
impurities. This increase at $T=0$  is real, can be huge and may
be observable, in principle. Since the effects of the
electron-electron interactions decrease with the elastic
scattering strength \cite{AAK}, the increasing electron mean free
path will actually {\em decrease} the effect of the electronic
Debye-Waller factor (\ref{DW1}) and help the vibrational one
(\ref{DW}) dominate. The analogy with the identification of the
decrease of quantum effects via the Debye-Waller factor as
"decoherence" would now suggest to define this increase as  {\em
"antidecoherence"} by a suitable coupling to zero-point modes of a
bath. It is hoped that by now the reader does not have to resist
the urge to rush to the word processor in order to produce an
announcement of this increase of quantum interference by the
coupling to the bath, as a revolutionary new physical phenomenon,
with a  range of important applications.

\section{V. Conclusions, real metals}
Obviously, the residual ($T \rightarrow 0$) resistivity of the
metal is determined by the elastic scattering  off the defects,
renormalized by a Debye-Waller type factor similar to the one
discussed above, including both the electron-phonon and
electron-electron interactions. All the low-temperature physics of
the metal is given in terms of this renormalized mean-free-path
(and mass). Strictly speaking, we have basically repeated and
explained the arguments of refs \cite{Schm-Berg}. These were
devised to refute the "zero temperature decoherence" by the
phonons. While we presented the detailed discussion for a single
"Einstein model" oscillator (which is an excellent test-case for
the theory), it is clear that the results are valid for an
arbitrary set of harmonic oscillators to which the conduction
electron is coupled, and for which the theorem about the gaussian
distribution we used is of course valid as well. Nothing of
principle will change in this argument if the phonon harmonic
oscillators are replaced by another set of {\em harmonic}
oscillators, such as those taken by Nyquist \cite{Nyquist},
Feynmann and Vernon \cite{FV}, Zwanzig \cite{Zwan} and Leggett and
Caldeira \cite{CL} to represent the dissipation in the solid.
Moreover, there is no need to use just an "ohmic" model. An almost
arbitrary set of oscillators can be used, including "subohmic",
"superohmic" or whatever one pleases (see below why we used the
word "almost"). These models are formally solvable as long as the
coupling to the oscillators is {\em linear}, and the solution is
elementary. In the same fashion as with the well-known fact that
the observed physics of an electron in vacuum automatically
includes the effect of the zero-point motion of the latter, the
observed physics of the electron in a real piece of material
automatically has all the renormalizations by whatever harmonic
modes of that metal that are coupled to the electron. Special low
energy excitations, such as TLS, magnetic modes, etc., can yield
nontrivial \cite{IFS} low-temperature decoherence for linear
response.

Of course, should there be special harmonic (or anharmonic) modes
with an anomalously  large DOS at low energies, the dephasing rate
due to them  would be anomalously large at the corresponding low
temperatures. It, as well as that by the TLS, should however
vanish at the strict $T \rightarrow 0$ limit, unless a massive
ground-state degeneracy or a pathologically increasing low energy
DOS would exist. We used here a scattering picture, as in the
Landauer description: an electron comes from the outside and its
transmission by the system determines the conductance of the
latter \cite{MW}. The Fermi-surface electron comes and exits at
the same energy. Its transmission amplitude across the system is
given by a sum of amplitudes, each of which multiplied by a
Debye-Waller type factor similar to Eqs. \ref{DW1},\ref{DW2}.
Since no excitations can be left by this electron at the $ T
\rightarrow 0$  metal whose ground state is non-degenerate, no
decoherence (in the sense of the interference terms becoming
smaller than the "classical" terms) can follow. This also applies
when the conventional low-temperature Fermi liquid picture is used
for the electrons in the  conductor. We found that unless further
special ingredients are introduced \cite{FLT}, this picture is
theoretically stable at low enough temperatures and energies. It
should be emphasized that such a scattering or a quasiparticle
picture is convenient in order to have a clear definition of
interference and decoherence. We did not mention the subtleties
associated with models in which the interaction with {\it the
same} environment  is always on. How to define the interfering
particle  becomes a nontrivial issue in that case. Such subtleties
are theoretically interesting but are hardly of much physical
relevance. When the electron moves in a real solid the
interactions are of finite range and hence it interacts with {\em
different} degrees of freedom when it is at different positions
\cite{Doron}. Therefore the Physics is similar to the scattering
situation \cite{CI} and not to models where the same interactions
are on all the time.

We repeat that we  did not rule out, and have not discussed,
special situations  where some additional new ingredient
\cite{FLT} will cause anomalous decoherence/relaxation for the
low-energy electrons; nor have we discussed the nontrivial
experimental situation in depth. Regardless of all that, the
statement that {\em decoherence is due to "inelastic" processes
(change of state of the environment)} is the valid guiding
principle for this problem. No ingenious model or advanced
technique can go around that.  Delving deeper into models that are
argued to violate that principle may only be useful in order to
expose their limitations or as diagnostics for the calculation. We
finally add that during the preparation of this note, the present
interpretation was conveyed to F. Guinea and found to be
completely consistent  with his model calculations \cite{Paco}.
These are therefore consistent with the notion that there is no
decoherence  in the $ T \rightarrow 0$ limit.

The deep question of how good is the description of  a real metal
via coupling to harmonic oscillators was avoided here. Besides not
distinguishing between resistance due to elastic or inelastic
scattering, that picture is for sure not exact and corrections may
well exist and be relevant \cite{Doron,Paco}. For example, in the
real system, there will be (hopefully small) corrections to the
gaussian approximation for the fluctuations \cite{Dima}, which
have not been  seriously treated either here or elsewhere.
Independently of that, the underlying statement that without
transfer of excitation there will be no dephasing, should still be
valid.

We  briefly  summarize the theoretical statements of this note.
The coupling of a particle that can not lose energy to
environmental degrees of freedom at $T \rightarrow 0$, can modify,
sometimes very seriously, the quantum behavior. This can go in
either direction and is {\em not} decoherence. On the other hand,
the experiments still point out to some nontrivial effects that
are not understood as yet,  in the low-temperature Physics of
disordered metals.
 \vspace{1cm}

{\bf Acknowledgements}\\

The author was fortunate to learn about the Debye-Waller factor
from Harry (Zvi) J. Lipkin and from the late Solly G. Cohen and
Israel Pelah and to collaborate on decoherence with Yakir Aharonov
and Ady Stern. Past discussions with Walter Kley and (later) with
Ted Schultz and collaboration with Leon Gunther and Benny Gavish
on these issues at low dimension have been very helpful as well.
Gerd Sch\"on is thanked for useful correspondence which helped
clarify the issue. More recent discussions with Michael Tinkham,
with  Richard Webb, with Doron Cohen and with Zvi Ovadyahu, whose
experiments have highlighted a very relevant issue which has been
posed already in Ref. \cite {Webb}), as well as with Uri Gavish
and Yehoshua Levinson, are gratefully acknowledged. The writing of
this  note began when the author was in the Nanoscience program at
the Institute for Theoretical Physics, Santa Barbara and was
supported there by the NSF Grant PHY99-07949. Work at the Weizmann
Institute was supported by a Center of Excellence of the Israel
Science Foundation, Jerusalem and by the  German Federal Ministry
of Education and Research (BMBF) within the Framework of the
German-Israeli Project Cooperation (DIP).


\begin{thebibliography}{99}

\bibitem{footnote1} We shall use here the terms "dephasing" and "decoherence"
interchangeably and will not even attempt to decide which of them
is more appropriate. The clearest definition of such a process is
provided by a two-wave interference situation where the intensity
is a sum of a classical term and an interference term, as in Eq.
\ref {TwoWave} below. Decoherence is defined via the "contrast" of
the observed picture, i.e. by the relative decrease of the latter
compared to the former.

\bibitem{SAI}
A. Stern, Y. Aharonov,  and Y. Imry,  Phys. Rev. {\bf A41}, 3436
(1990) ; and in  G. Kramer, ed. {\it Quantum Coherence in
Mesoscopic Systems}, NATO ASI Series no. 254, Plenum., p. 99
(1991).

\bibitem{FV} R. P. Feynmann and F. L. Vernon  Jr., Ann. Phys. (N.Y.) 24, 118 (1963).


\bibitem{book}
Y. Imry, {\it Introduction to Mesoscopic Physics}, Oxford
Unversity Press (1997, 2nd edition  2002).

\bibitem{Kumar}
N. Kumar, D. V. Baxter, R. Richter and J. O. Strom-olsen Phys.
Rev. Lett. {\bf 59}, 1853 (1987).

\bibitem{Schm-Berg} J. Rammer, A. L. Shelankov and
A. Schmid, Phys. Rev. Lett. {\bf 60}, 1985 (1988); G. Bergmann,
Phys. Rev. Lett. {\bf 60}, 1986 (1988).




\bibitem{Webb} P. Mohanty, E.M. Jariwala and R.A. Webb,
Phys. Rev. Lett. {\bf 78}, 3366 (1997); P. Mohanty and R.A. Webb,
Phys. Rev. {\bf B55}, 13452 (1997); R.A. Webb, P. Mohanty and E.M.
Jariwala in {\it Quantum Coherence and Decoherence}, proceedings
of ISQM, Tokyo (1998), K. Fujikawa and Y.  A. Ono, eds. North
Holland, Amsterdam (2000).

\bibitem{Berg} G. Bergmann, Phys. Reports {\bf 107}, 1 (1984),
an excellent review of  weak-localization.

\bibitem{AAK}
B.L. Altshuler, A.G. Aronov,  and D.E. Khmelnitskii, J. Phys. {\bf
C15}, 7367 (1982) .





\bibitem{Zvi} Besides using Fermions, there are other ways to block the inelastic
scattering from  $T \rightarrow 0$ systems. A simple and
particularly transparent one was suggested to the author by H. J.
Lipkin, private communication. The lowest excited state of the
scatterer can be taken to have an excitation energy  larger than
the initial energy of the scattered particle.

\bibitem{footnote2} Actually, in more sophisticated descriptions, the linear
transport coefficient (e.g conductance) is described via {\em
equilibrium} properties (i.e. dynamic correlation functions) of
the system. Hence, there is absolutely no need to invoke
nonequilibrium Physics.

\bibitem{CI} D. Cohen and Y. Imry, Phys. Rev. {\bf B59}, 11143
(1999).


\bibitem{isqm} Y. Imry,
in {\it Quantum Coherence and Decoherence}, proceedings of ISQM,
Tokyo (1998), K. Fujikawa and Y.  A. Ono, eds. North Holland,
Amsterdam (2000).



\bibitem{IFS} Y. Imry, H. Fukuyama and P. Schwab,
Europhys. Lett, {\bf 47}, 608 (1999).

\bibitem{kha} Yu. B. Khavin, M. E. Gershenson and A. L. Bogdanov,
Phys. Rev. Lett., {\bf 81}, 1066 (1998), Phys. Rev. {\bf B58},
8009 (1998); M. E. Gershenson et al., Sov. Phys. Uspekhi {\bf 41}
(2), 186 (1998).




\bibitem{Norm}
A. B. Gougam, F. Pierre, H. Pothier,  D. Esteve and N. O. Birge,
J. Low Temp. Phys. {\bf 118}, 447 (2000); see also: F. Pierre, H.
Pothier,  D. Esteve and M. H. Devoret, ibid.; F. Pierre, Ph.D.
thesis, Universit\'e Paris VI (2000), unpublished.


\bibitem{Ov00} Z. Ovadyahu, Phys. Rev. {\bf B63}, 235403 (2001).


\bibitem{zaikin}
D.S. Golubev and A.D. Zaikin, Phys. Rev. Lett., {\bf 81}, 1074
(1998); Phys. Rev. {\bf B59}, 9195 (1999); Phys. Rev. Lett., {\bf
82}, 3191 (1999); cond-mat/0103166.

\bibitem{ale} I. L. Aleiner, B. L. Altshuler, M. E. Gershenson,
Waves in Random Media {\bf 9}, 201 (1999); Phys. Rev. Lett., {\bf
82}, 3190 (1999).
\bibitem{alt} B. L. Altshuler, M. E. Gershenson, I. L. Aleiner,
Physica {\bf A3}, 58 (1998).

\bibitem{vonDelft} V. Ambegaokar and J. von Delft, private
communication, see:
http://www.theorie.physik.uni-muenchen.de/~vondelft/dephasing.


\bibitem{Schoen} D.S. Golubev, A.D. Zaikin, Gerd Sch\"on {\it On
Low-Temperature Dephasing by Electron-Electron Interaction},
cond-mat/0110495, to be published in J. Low Temp. Phys.

\bibitem{KB} T.R. Kirkpatrick and D. Belitz,  {\it Absence of electron
dephasing at zero temperature}, cond-mat/0111398; Dmitri S.
Golubev, Andrei D. Zaikin, Gerd Sch\"on {\it Comment on "Absence
of electron dephasing at zero temperature"}, cond-mat/0111527;
T.R. Kirkpatrick and D. Belitz, {\it Reply to comment on "Absence
of electron dephasing at zero temperature"}, cond-mat/0112063.




\bibitem{But}
P. Cedraschi, V. V. Ponomarenko and M. B\"uttiker, Phys. Rev.
Lett. {\bf 84}, 346 (2000) and Ann. Phys. (NY), in press.

\bibitem{Nyquist} H. Nyquist, Phys. Rev. {\bf 32}, 110 (1928).

\bibitem{Zwan} R. Zwanzig, J. Stat. Phys. {\bf 9}, 215 (1973).

\bibitem{CL} A. O. Caldeira and A. O. Leggett,  Ann. Phys. (N.Y.) {\bf 149} 374; 153, 445(E) (1984).

\bibitem{Doron} D. Cohen J. Phys. {\bf A31}, 8199 (1998);
Phys. Rev. {\bf E55}, 1422 (1997); Phys. Rev. Lett. {\bf 78}, 2878
(1997). cond-mat/0103279 , to be published in Phys. Rev. E.


\bibitem{VH}
L. van Hove, Phys. Rev {\bf 95}, 249 (1954). In this classic
paper, the dynamic correlation function for the density $\langle
\hat{n}_{q}(0) \hat{n}_{-q}(t) \rangle$, was introduced, it was
shown that its time Fourier transform, $S(q, \omega)$, yields the
inelastic Born scattering cross section from the system, the
properties of the latter were analyzed and the connection to the
dissipative response indicated.

\bibitem{Kittel}
C. Kittel, {\it Quantum Theory of Solids} Wiley (1963).

\bibitem{Messiah}
A. Messiah, {\it Quantum Mechanics}, North Holland, Amsterdam,
1961.

\bibitem{stat}
This is, stricly speaking, valid under the proviso that adding to
the scattered particle a few quanta of the harmonic oscillator
does not change its wavevector significantly, this has been termed
by van Hove "the static approximation", very well valid for x-ray
scattering, but not, for example, for slow neutron scattering in
solids. This is however a technical detail which is irrelevant to
our discussion

\bibitem{Zemach} A. C. Zemach and R. J. Glauber Phys. Rev. {\bf 101}, 118,129
(1956).

\bibitem{Lipkin} H.J. Lipkin, {\it Quantum Mechanics}, North-Holland,
Amsterdam, 3rd printing (1992).


\bibitem{Fermi} E. Fermi, Ricerca Scientifica {\bf 7}, 13 (1936).

\bibitem{multip} One might do better and include, in principle,  the
interscatterer multiple scattering, by using a pseudopotential
specific to the given structure. This is however totally
irrelevant to our discussion here. We only mention it because of
the misuse of the term "nonperturbative" by some of the
participants in the current controversy.



\bibitem{PE} See e.g. G.-L. Ingold and Yu. V. Nazarov, in {\it Single Charge
Tunneling}, eds. H. Grabert and M. Devoret, Plenum 1992, Chapter
2.





\bibitem{GI} Y. Imry and L. Gunther,
Phys. Rev. {\bf B3}, 3939 (1971).

\bibitem{IG}  Y. Imry and B. Gavish, J. Chem. Phys. {\bf 61}, 1554
(1974), considered a  combination of 1D phonons and dissipation,
in the classical limit.



\bibitem{Holstein} T. Holstein,   Ann. Phys. (N.Y.) {\bf 8}, 343 (1959).

\bibitem{Schmid} A. Schmid, Phys. Rev. Lett. 51, 1506–1509 (1983)

\bibitem{Legg} A. J. Leggett et al., Rev. Mod. Phys. {\bf 59},
1 (1987).

\bibitem{Stern} A. Stern, Ph.D. Thesis, Tel-Aviv University (1990),
unpublished.


\bibitem{expl}
The Feynmann-Vernon--Caldeira-Leggett picture models the finite
low-frequency mobility, $ \mu \equiv \mu(\omega\rightarrow 0)$, of
a particle whose velocity operator is ${\hat v}$,  by coupling it
linearly to a continuum of harmonic oscillators. The coupling to
the oscillator having a very low frequency $\omega_\lambda$ and
position operator ${\hat x_\lambda}$, is taken to be $c_\lambda
\hat v \hat x_\lambda~.$ This looks like a coupling to a vector
potential $\sum_\lambda c_\lambda \hat x_\lambda$. For a quick and
dirty derivation one may follow the Onsager-type idea of using the
macroscopic laws to describe the behavior of the fluctuations. We
thus write the fluctuating $v$ as the linear response to the
fluctuating instantaneous value of the driving field, taken, in
the semiclassical case, to be a c-number. Expressing the electric
field as the time-derivative of the vector potential, it follows
that the power spectrum of ${\hat v}$ is proportional to $\mu^2
\sum_\lambda \omega_\lambda^2 |c_{\omega_\lambda}|^2
(x_{\lambda})^2_\omega$, where $(x_{\lambda})^2_\omega$ is the
power spectrum of $x_{\lambda}$. We remember that the latter is
given by $\frac{\hbar}{M\omega}[ \delta(\omega + \omega_\lambda)
(\nu + 1/2) + \delta(\omega - \omega_\lambda)\nu] $. Here $\nu$ is
the average number of excited oscillator quanta $[exp (\hbar
\omega / k_B T) - 1]^{-1}$. The fluctuation-dissipation theorem
\cite{Kubo} relates \cite{Moriond} $\mu$ with the (negative
frequency at $T \rightarrow 0$, and in general the difference
betwen the negative and positive frequency branches of the)
power-spectrum of ${\hat v}$, divided by $\omega$. It follows that
$$\sum|c_{\lambda}|^2 \delta(\omega - \omega_\lambda)\propto \frac{M}{ \hbar \mu} ~.$$
If the coupling is written as: $\sum_\lambda C_\lambda \hat x \hat
x_\lambda$, then the coefficients are related by $|C_\lambda|^2 =
|c_\lambda|^2  \omega ^2$. The usual explanation of this model
goes by noting that the excitation of these low-frequency
oscillators provides the viscous force on the slow particle. The
(related) understanding employing the fluctuation-dissipation
theorem, pictures the particle as coupled to a set of oscillators
designed to enable the system to have a low-frequency absorption
mimicking that of a diffusing particle.


\bibitem{Im} Actually, in the quantum regime the correlator
in the time domain develops an imaginary part due to the
noncommutativity of the exponent in the operator $\exp[i \phi(t)]$
with itself at two different times. This results in an asymmetry
in the frequency domain \cite{Uri}. As a result, a  term of the
type $i \sin\omega t$ is added \cite{Kittel,PE} to the $\cos\omega
t$ in Eq. \ref{DW1}. At $T = 0$ this converts Eq. \ref{DW1} to:

\bq \exp[- const' \int_{-\infty}^\infty d\omega \eta(\omega)\,
(\frac{1-\exp i\omega t}{\omega})], \label{DWright} \ee

\noindent which can be seen to eliminate the $\omega < 0$ portion
of the power spectrum of $\exp[i \phi(t)]$. This portion yields
the energy transfer from the oscillators to the particle. Such a
transfer is ruled out when the oscillators are at  $T = 0$. At
finite temperatures this imaginary part is what brings about the
correct (detailed-balance type) ratio of the energy-gain and
energy-loss branches of the power spectrum. It is not crucial,
however, for the Debye-waller factor, which is the focus of our
discussion here.

\bibitem{Uri} U. Gavish, Y. Levinson, and Y. Imry,
{\it Detection of quantum noise}, Phys. Rev. {\bf B RC}, R10637
(2000).


\bibitem{Kubo}   R. Kubo,  J. Phys. Soc. Japan {\bf 12}, 570 (1957);
{\bf 17}, 975 (1962).



\bibitem{Moriond}
U. Gavish, Y. Levinson and Y. Imry {\it Quantum noise, detailed
balance and Kubo formula in non-equilibrium quantum systems}, to
be published in the Proceedings of the 2001 Rencontres de Moriond:
{\it Electronic Correlations: from Meso- to Nanophysics}, T.
Martin, G. Montambaux and J. Tr$\hat a$n Thanh V$\hat a$n eds.
EDPSciences 2001.

\bibitem{Kagan}
Yu. Kagan and A. P. Zhernov, Zh. Eksp. Teor. Fiz. 50, 1107 (1966)
[Sov. Phys. JETP 23, 737 (1966)].

\bibitem{Misha} P. M. Echternach, M. E. Gershenson and H. M. Bozler, Phys. Rev.
B 47, 13659 (1993).


\bibitem{MW} It should perhaps be emphasized that in spite of
claims otherwise, the Landauer (two-terminal) formulation is
equivalent to the linear response theory for  the same geometry,
at least at $T \rightarrow 0$. We have in mind a finite sample,
with arbitrary interactions, connected by ideal and noninteracting
leads to particle reservoirs. The Landauer conductance is defined
by the transmission of electrons across this sample. The linear
response conductance can be evaluated for the same model. Their
equivalence as $T \rightarrow 0$ for an interacting system was
proven by Y. Meir and N. Wingreen, Phys. Rev. lett. {\bf 68}, 2512
(1992) [See also: T. K. Ng and P. A. Lee, Phys. Rev. Lett. {\bf
61}, 1768 (1988)]. The case of finite temperatures where inelastic
scattering may occur is more complicated, but it is not relevant
to the present discussion, which concerns the $T \rightarrow 0$
(no inelastic scattering) limit only. This finite temperature case
will hopefully be treated in a future publication.



\bibitem{FLT} In situations with such anomalous low-energy
behavior, the Fermi liquid theory will break down at low
temperatures. Two canonical examples for that are the two-channel
Kondo model and the Luttinger liquid. It is perhaps helpful to
point out, however, that in the real world, there may exist
factors that can destabilize these deviations from the FLT at low
enough temperatures (such as as a small interchannel coupling or a
finite separation between the two states \cite{Zawa} in the
two-channel Kondo model, a finite length of the Luttinger liquid).
Likewise, disordered systems that have theoretically a
ground-state degeneracy (for example, the ice model \cite
{Pauling} or glasses) may have a small real-life splitting of that
degeneracy or never experience it within  any finite measurement
time. On the other hand, situations where truly robust deviations
from the FLT as $T \rightarrow 0$ exist are obviously of genouine
interest.

\bibitem{Zawa} A. Zawadovski, Jan von Delft and D. C. Ralph,
Phys. Rev. Lett. {\bf 83}, 2632 (1999); see also: I. L. Aleiner,
B. L. Altshuler, Y. M. Galperin and T. A. Shutenko {\bf 86}, 2629
(2001), cond-mat/0007430.

\bibitem{Pauling} L. Pauling, {\it The Nature of the Chemical Bond},
Cornell University Press, Ithaca, NY (1960), 3rd edn.; G.C.
Pimental and A.L. McClellan,  {\it The Hydrogen Bond}, Freeman,
San Francisco (1960).

\bibitem{Paco} F. Guinea, {\it Aharonov-Bohm oscillations of a
particle coupled to dissipative environment}. cond-mat/0112099,
(2001).

\bibitem{Dima} It has been pointed out to the author by D. E.
Khmel'nitskii that the treatments of Refs. \cite {AAK} and \cite
{SAI} are valid only within the gaussian approximation for the
fluctuations.









\end{thebibliography}
\end{document}